\documentclass[aps,floats,prl,showpacs,twocolumn ]{revtex4-1}

\usepackage{amsfonts,amsmath,mathrsfs} \usepackage{bm} \usepackage{dcolumn}
\usepackage{epsfig} \usepackage{latexsym}

\begin{document}

\title{Quantum phase transitions in the quasi-periodic kicked rotor}

\author{Chushun Tian, Alexander Altland, and Markus Garst}

\affiliation{Institut f{\"u}r Theoretische Physik, Universit{\"a}t
zu K{\"o}ln, K{\"o}ln, 50937, Germany}


\begin{abstract}
  We present a microscopic theory of transport in quasi-periodically
  driven environments (`kicked rotors'), as realized in recent atom
  optic experiments. We find that the behavior of these systems
  depends sensitively on the value of Planck's constant $\tilde h$:
  for irrational values of $\tilde h/(4\pi)$ they fall into the
  universality class of disordered electronic systems and we derive
  the microscopic theory of the ensuing localization phenomena.  In
  contrast, for rational values the rotor-Anderson insulator acquires
  an infinite (static) conductivity and turns into a
  `super-metal'. Signatures of the corresponding metal/super-metal
  transition are discussed.
\end{abstract}


\maketitle

The quantum kicked rotor (QKR) is one of the most prominent model
systems of quantum nonlinear dynamics (quantum
chaos)~\cite{Casati79}. It comprises a one--dimensional quantum
particle subject to periodic boundary conditions and
sequential driving in time. In spite of the nominal
simplicity of this system
--- the QKR Hamiltonian is defined in terms of only two dimensionless
parameters, kicking strength, $K$, and time in relation to
(effective) Planck's constant, $\tilde h$ \cite{unit} --- it
displays striking parallels to disordered electronic systems
\cite{Fishman83,Altland10}, and the ensuing connections to condensed
matter physics have been a subject of fundamental research for more
than three decades (cf. Ref.~\cite{Altland10} and references
therein). The general interest in the QKR took a leap forward when
the system was experimentally realized in a cold atom
setting~\cite{Raizen95}, and localization phenomena otherwise
predicted for disordered multi-channel quantum wires were observed.
This has been the first observation of strong Anderson localization
in a quasi-one dimensional setting (although the full microscopic
correspondence to disordered quantum wires was established only
recently~\cite{Altland10}.) The next experimental breakthrough
occurred in 2008 when an effectively higher dimensional rotor, the
so-called quasi-periodic kicked rotor, was realized in a gas of cold
cesium atoms, and an Anderson \textit{transition} was
observed~\cite{Deland08}. By now, even signatures of the critical
states emerging at the transition point have been
seen~\cite{Deland10}, and it stands to reason that the
quasi-periodic QKR makes for an almost ideal environment to study
Anderson type critical phenomena.

At the same time, the correspondence between the quasi-periodic QKR
and $(d>1)$--dimensional disordered systems is not as well
understood as in the $d=1$ case, and this is a gap which we aim to
close in this Letter. In the quasi-periodic QKR, $d$--dimensional
behavior is simulated by modulated driving at $d$ different
frequencies. Below, we will map the low energy physics of this
system onto an effective field theory equivalent to the non-linear
$\sigma$-model of disordered metallic systems \cite{Efetov97}. For
generic values of the system parameters, this construction
establishes the connections to disordered systems, and it explains
the observation of Anderson type criticality. However, the rotor is
not a genuine metal, and these differences show in anomalies at
certain (non-generic) parameter values. Specifically, the
one-dimensional QKR displays so-called quantum resonances (cf.
Ref.~\cite{Altland10} and references therein) at rational values
$\tilde h=4\pi\, p/q$, $p,q\in \Bbb{N}$ coprime. At resonance, the
system behaves like a quasi-one dimensional \textit{ring} (in the
space of rotor-angular momenta) of radius $q$, and no Anderson
localization is observed.

\begin{figure}[h]
  \centering
\includegraphics[width=8cm]{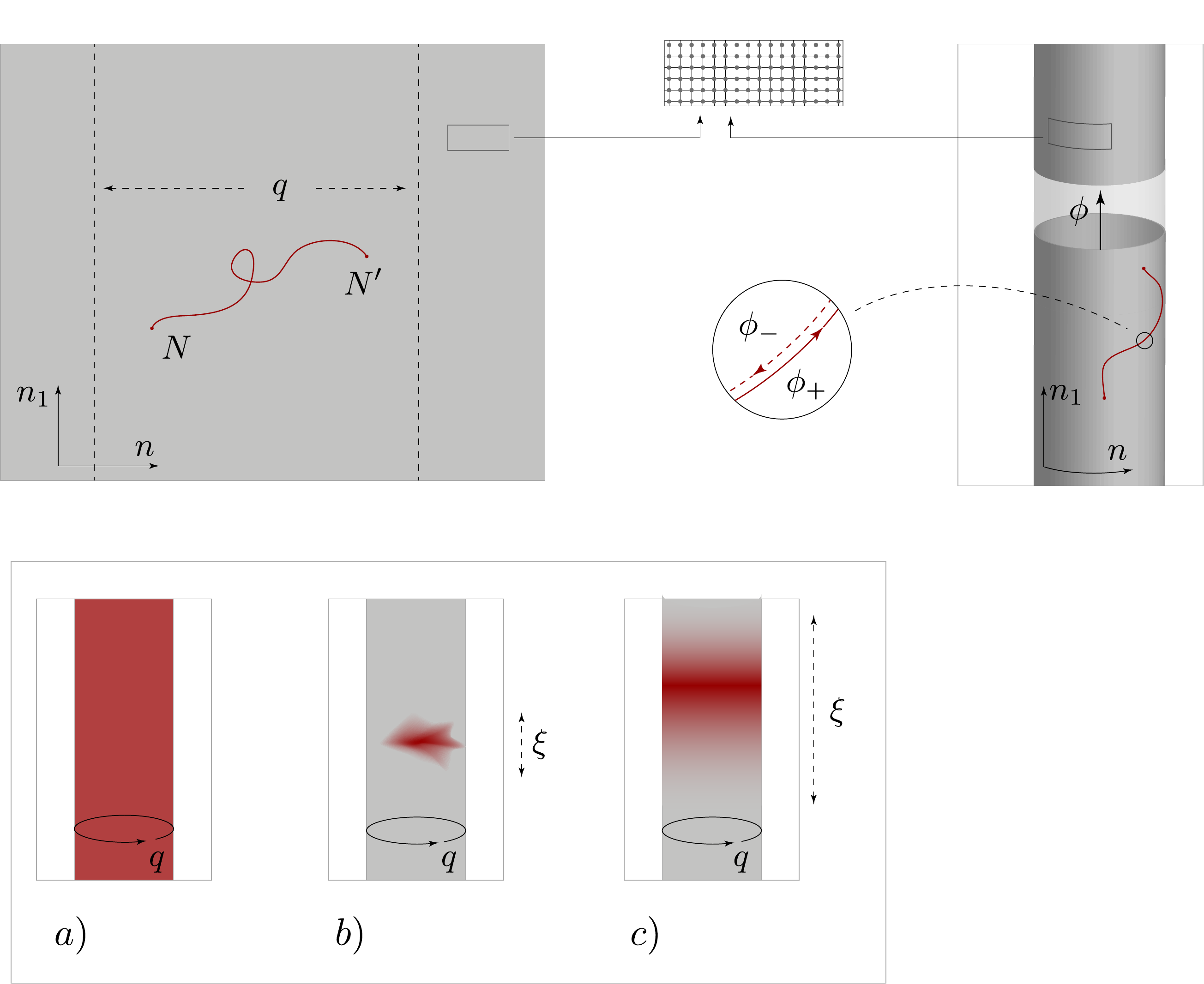}
\caption{The $d=2$ QKR at resonance.  Left: $q$-periodic angular
  momentum space. Right: compactification to cylinder of circumference
  $q$, threaded by flux $\phi$. Grey shaded areas represent a $2d$
  lattice of angular momentum sites. Transport between sites $N$ and
  $N'$ takes place through phase coherent propagation of advanced
  (solid) and retarded (dashed) quantum amplitudes subject to
  different flux values $\phi_\pm$. Bottom, the three distinct regimes
  in virtual directions, a) unbound diffusion, b) localization at
  $\xi<q$, and c) localization at $\xi>q$.}
  \label{fig}
\end{figure}

Below we will show that in dimensions $d>1$ the system displays even
richer behavior. At $\tilde h = 4\pi p/q$ the system
\textit{compactifies}, e.g., for $d=2$ a nominally two dimensional
QKR maps onto the surface of an infinitely long cylinder of finite
circumference $q$ (cf. Fig.~\ref{fig}, top). As far as bulk
localization phenomena are concerned, this entails a dimensional
reduction to quasi $d-1$ dimensions (localization along the cylinder
axis). However, measurable response functions dominantly couple to
its compact dimension, and this has a number of interesting
consequences: below we consider a (measurable) observable which, in
the light of the above analogies to metallic system, plays a role
analogous to an optical conductivity. We find that in the quasi
$(d-1)$-dimensional localized regime, the system actually shows
`super-metallic' behavior (diverging conductivity). In contrast, for
quasi $(d-1)$-dimensional metallic regimes (e.g., the quasi
three-dimensional metallic phase realized by driving at four
different frequencies), the conductivity remains finite. For $d>3$,
an Anderson transition in quasi $(d-1)$-dimensions may be driven by
changing the kicking strength, and this manifests itself in a
metal/super-metal transition in observable quantities. A survey of
the three different phases realizable in the QKR is shown in table
\ref{tab:1}.

\begin{table}[h]
  \centering
    \begin{tabular}{|l||l|l|}\hline
properties in quasi & non-generic Planck's & generic Planck's \cr
($d-1$)-dimensions & constant & constant \cr \hline\hline localized
&\textit{super-metallic},& \textit{insulating}, \cr & infinite
conductivity& zero conductivity \cr\hline metallic &
\textit{metallic}, ($d\ge 4$)& \textit{metallic}, ($d\ge 3$)\cr &
finite conductivity & finite conductivity \cr\hline
   \end{tabular}
  \caption{Phases realizable in the quasi-periodic
  QKR}
  \label{tab:1}
\end{table}

In dimensionless units \cite{unit}, the time dependent Hamiltonian
of the system is defined as ${\hat H} (t) = \frac{1}{2} {\tilde
h}^2\hat n^2 +KV(t) \sum_m\delta (t-m)$. Here,
$\theta$ is the rotor's angular variable, $\hat
n=-i\partial_{\theta}$ the angular
momentum operator, and $ V(t)\equiv V(\cos\theta,\cos(\theta_1+\omega_1
t),..., \cos(\theta_{d-1}+\omega_{d-1} t)), $ the kicking potential.
The $(d-1)$ frequencies, $\omega_{1},...,\omega_{d-1}$, are
incommensurate to the kicking frequency $2\pi$ and among themselves,
and $\theta_{1},...,\theta_{d-1}$ are $d-1$ arbitrary phases. We
assume $V$ to be symmetric with respect to its $d$ arguments, and of
unit characteristic variation.
Later we will see that the detailed form of $V$
determines the diffusion constant of the system, but  is largely
inessential otherwise. We assume $\tilde
h/4\pi = p/q
$
where the limit of an irrational value may be taken by sending $p,q\to \infty$.

As in the experimental applications, we consider the evolution of a wave
function initially uniform in $\theta$. The localization properties of
such states are probed by the
correlation function
\begin{align}
  \label{eq:9}
  E(t) \equiv
\frac{1}{2}\sum_{n} \overline{|\langle n|\Pi_{m=1}^t \hat
U'(m)|0\rangle|^2}\, n^2
\end{align}
where $\hat U'(m)\equiv \exp(i{\tilde h \over 2}\hat n^2)
\exp(i\frac{K}{\tilde h} \hat V(m))$ is the Floquet operator and the
overline stands for the average over the parameters $\theta_i$.
The mapping to an effectively higher dimensional
system~\cite{Casati89} is achieved by interpreting $|\Theta\rangle\equiv
|\theta,\theta_1,\dots,\theta_{d-1}\rangle$ as a
$d$-dimensional coordinate vector, comprising a `real' angular
coordinate, and a generalization of the parameters $\theta_i$ to
`virtual' coordinates. Similarly, we introduce a
$d$-dimensional angular momentum state,
$|N\rangle=|n,n_1,\dots,n_{d-1}\rangle$ where $\hat n_i =
-i\partial_{\theta_i}$ is conjugate to $\theta_i$, with eigenvalues
$n_i\in \Bbb{Z}$. Defining the operator $\hat \Phi(m) \equiv
\exp(-im\sum_i \omega_i \hat n_i)$, it is then not difficult to
verify that the `gauge transformed' Floquet operator $\hat U(m)
\equiv \hat \Phi(m+1) \hat U'(m) \hat \Phi^{-1}(m)$ becomes
time-independent, viz. $\hat U \equiv \hat T(\hat N) \hat W(\hat
\Theta)$, $\hat T(\hat N) \equiv e^{i{\tilde h\over 2} \hat
n^2+i \sum_i\omega_i
    {\hat n}_i}$, $\hat W(\hat \Theta)\equiv
\exp[\frac{iK}{\tilde
h}V(\cos\theta,\cos\theta_1,\cdots,\cos\theta_{d-1})]$, and that
$E(t) = \frac{1}{2} \sum_{N} |\langle N |\hat U^t|0\rangle|^2 n^2$.
We have, thus, traded the time dependence of the
original problem for an effective extension to a multi-dimensional Hilbert space spanned by the states $|N\rangle$.

The effective Floquet operator $\hat U$ possesses two fundamental
symmetries: time reversal symmetry
\cite{Smilansky92} $\mathrm{T}: t\rightarrow -t, \hat
\Theta\rightarrow -\hat \Theta$, and invariance under the translation ${\hat
n}\rightarrow {\hat n}+ q$. Exploiting the latter, it
is straightforward to verify that the variable $E(t)$ affords the
representation
\begin{align}
\label{eq:8}
  E(t) = {q\over 2} \int\limits_0^{2\pi/q} \frac{d\phi}{2\pi}  \partial^2_{\phi_+\phi_-}
\mathrm{tr}\left(   \hat U^t_{\phi_+}\, \delta_{\hat N 0}\, \hat
U^{t\dagger}_{\phi_-}\right) \big|_{\phi_\pm=\phi},
\end{align}
where `tr' is a
trace over all states $|N\rangle$
whose real coordinate $n\in \{0,\dots ,q\}$ is restricted to a
compact `unit cell', and we have defined the `Bloch-Floquet'
operator $\hat U_\phi \equiv \hat T(\hat N) \hat W( \hat
\Theta+\phi)$, where $\hat \Theta +\phi \equiv (\hat
\theta+\phi,\theta_1,\dots, \theta_{d-1})$. As in the quantum
mechanics of periodic structures, the summation over
a Bloch phase, $\phi$, enables us to
compactify $n$-space to a ring of circumference $q$ with periodic
boundary conditions. The shift of the angular variable $\hat \theta
\to \hat \theta+\phi$ shows that $\phi$ couples to the system as an
Aharonov-Bohm flux, cf. Fig.~\ref{fig}, top.

Eqs. (\ref{eq:9}) and (\ref{eq:8}) are different (yet equivalent)
ways of probing the time dependent spreading of angular momentum
states. Anticipating a competition of classical diffusion and
quantum localization, we expect three qualitatively distinct cases
(cf. Fig.~\ref{fig}, bottom): if the localization length, $\xi$ is
infinitely large, unbound diffusive spreading $n^2\sim Dt$
characterized by a diffusion coefficient, $D$, leads to a linear
increase $E(t) \sim Dt$ (\textit{metallic regime}). In contrast, for
$\xi<q$ we expect saturation, $E(t\gg \xi^2/D) \sim \mathrm{const.}$
(\textit{localized regime}). (A finite size correction $\sim t^2$
exponentially small in $e^{-q/\xi}$ will be ignored.)  Finally, in
cases where $\xi>q$, the system behaves similar to a finite quantum
system of characteristic quasilevel spacing $
\sim 1/(q\xi^{d-1})$. For large times, $t\gg q\xi^{(d-1)}$,
individual states of this system can be resolved and a formal
decomposition of $\hat U$ in quasi-energy states shows that $E(t
) \sim t^2$ (\textit{super-metallic regime}).

To quantitatively describe these regimes, we define the resolvent
operators $\hat G^\pm_\phi(\omega_\pm) \equiv (1-(e^{i\omega_\pm}
\hat U_\phi)^{\pm 1})^{-1}$, where $\omega_\pm \equiv \omega_0\pm
{1\over 2}(\omega+i0)$. The Fourier transform $E(\omega) =
\int_0^\infty dt\, e^{i \omega t} E(t)$, then assumes the form $
E(\omega) =\int_0^{2\pi} \frac{d\phi}{2\pi}
\partial^2_{\phi_+\phi_-}\big|_{\phi_\pm=\phi} Y(\phi_+,\phi_-,\omega)$,
where $Y(\phi_+,\phi_-,\omega)\equiv \left\langle \mathrm{tr}(
  \hat G^+_{\phi_+}(\omega_+)\,\delta_{\hat N 0}\,
  \hat G^-_{\phi_-}(\omega_-))\right\rangle_{\omega_0}$ and
$\langle \dots \rangle_{\omega_0} = \int_0^{2\pi}
\frac{d\omega_0}{2\pi}$. Apart from an overall factor $\omega$, this
resembles the two-particle response function employed to compute the
optical conduction properties of electronic
systems.

To make further progress, we describe the correlation function $Y$
in terms of a low energy effective field theory. The technical
details of this mapping \cite{at} are nearly identical to those
of our earlier treatment of the one-dimensional
rotor~\cite{Altland10}, and we here restrict ourselves to a brief
sketch of the principal steps. We start from a representation
of the correlation function $Y$
in terms of Gaussian
integral over superfields~\cite{Efetov97}:
\begin{align*}
 Y(\phi_+,\phi_-,\omega)= \int dN \int D(\bar \psi,\psi) \,
  \left\langle e^{-\bar \psi
      G^{-1}
      \psi}\right\rangle_{\omega_0}X[\bar \psi,\psi]\,.
\end{align*}
Here, the superfield $\psi=\{\psi_{N,\lambda,\alpha}\}$ where
$\alpha=\mathrm{b,f}$ distinguishes between commuting and
anti-commuting components, and $\lambda=\pm$ between retarded and
advanced components. The pre-exponential term is given by $X[\bar
\psi,\psi]=\psi_{N,+,\mathrm{b}}\bar
\psi_{0,+,\mathrm{b}}\psi_{0,-,\mathrm{b}}\bar \psi_{N,-,\mathrm{b}}
$ and $G^{-1}
= \mathrm{diag}\left(G^{-1}_{\phi_+}(\omega_+),
  G^{-1}_{\phi_-}(\omega_-) \right)$ is a matrix block-diagonal in
advanced/retarded space. To make progress with this expression, we
apply the color-flavor transformation \cite{Zirnbauer96}, an
integral transform that trades the integral over $\psi$ and
$\omega_0$ for the integration over an auxiliary field, $Z$:
$K(\phi,\omega)= \int D(Z,\tilde Z)\,(\dots)\exp(-S[Z,\tilde Z])$,
where
\begin{align*}
 S[Z,\tilde Z]=-\mathrm{str \,ln}(1-Z\tilde Z)+\mathrm{str\, ln}(1-
 e^{i\omega} \hat U^\dagger_{\phi_-} Z \hat U_{\phi_+} \tilde Z),
\end{align*}
`str' is the supertrace~\cite{Efetov97}, and we have temporarily
suppressed the pre-exponential terms for notational simplicity.
Here, $Z=\{Z_{N\alpha,N'\alpha'}\}$ is a bi-local supermatrix field,
subject to the constraints $\tilde Z_{\mathrm{b,b}}=
Z_{\mathrm{b,b}}^\dagger$, $\tilde Z_{\mathrm{f,f}}=
-Z_{\mathrm{f,f}}^\dagger$ and $|Z_{\mathrm{b,b}}
Z_{\mathrm{b,b}}^\dagger|<1$. The anti-commuting blocks
$Z_{\alpha,\alpha'}$ and $\tilde Z_{\alpha,\alpha'}$,
$\alpha\not=\alpha'$ are independent. Physically (cf.
Ref.~\cite{Altland10} for a more extensive discussion), the field
$Z_{N,\alpha;N',\alpha'} \sim \psi_{N,\alpha,+}\bar
\psi_{N',\alpha',-}$ describes the pair propagation of a retarded
and an advanced single particle amplitude at a slight difference in
frequency, $\omega$, and Aharonov-Bohm flux $ \varphi \equiv
\phi_+-\phi_-$.  The structure of the action $S[Z,\tilde Z]$ shows
that at these values field configurations $\hat U^\dagger_{\phi_-} Z
\hat U_{\phi_+} \sim Z$ near--stationary under the adjoint action of
the Bloch-Floquet operator dominantly contribute to the field
integral. The identification of these `slow modes' is facilitated by
passing to a Wigner representation, $Z_{N_1,N_2}\to Z_{N,\Phi}$,
where $N=(N_1+N_2)/2$, and $\Phi$ is dynamically conjugate to $N$.
Due to the fast relaxation of the dynamics in the space of angular
variables, $\Phi$, the modes of lowest action $Z_{N,\Phi}=Z_N$
depend only on $N$. These angular zero modes then produce the low
energy representation
\begin{align}
\label{eq:12} &  Y(\varphi,\omega) = -\int dN \int dQ \,e^{-S[Q]}
  \,\mathrm{str}(Q_NP) \mathrm{str}(Q_{0}\bar P),\nonumber\\
& S[Q] = -{1\over 8}\int dN\, \mathrm{str}
  (D(\partial_\varphi Q )^2 + 2i\omega \,Q\sigma^3_\mathrm{AR}).
\end{align}
Here, $P=E^{12}_\mathrm{AR}\otimes E^{11}_\mathrm{BF}$ and $\bar
P=E^{21}_\mathrm{AR}\otimes E^{11}_\mathrm{BF}$, where  the $2\times
2$-matrices $E^{ij}_\mathrm{AR/BF}$ act in the space of
$\lambda/\alpha$--indices, carry a unity at position $(i,j)$, and
zero elsewhere. The matrix field $Q$ is given by
\begin{align*}
  Q= \left(
    \begin{array}{cc}
      1 & Z \\
      \tilde Z & 1 \\
    \end{array}
  \right) \sigma_{\rm AR}^3 \left(
    \begin{array}{cc}
      1 & Z \\
      \tilde Z & 1 \\
    \end{array}
  \right)^{-1},
\end{align*}
where $\sigma^3_\mathrm{AR}=E^{11}_\mathrm{AR}-E^{22}_\mathrm{AR}$.
The fluctuations of these fields are governed by an action $S[Q]$
identical to the action of the `diffusive' nonlinear $\sigma$-model
of disordered metals (subject to an Aharonov-Bohm flux).
The rotor's diffusion constant is given by $D=
\frac{C}{2}(\frac{K}{\tilde h})^2$, where
$C= \overline{(\sin\theta\partial_x|_{x=\cos\theta} V)^2}$ is a
constant of ${\cal O}(1)$ whose detailed value depends on the
kicking potential. (The perturbative integration over non-zero mode
configurations $Z_{N,\Phi}$ generates corrections to $D$ of ${\cal
  O}(1/K)$ which we do not discuss here.) Finally, $\partial_\varphi
\equiv(\partial_n + i\varphi
[\sigma^3_\mathrm{AR},\;],\partial_{n_1},\dots \partial_{n_{d-1}})$,
is a covariant derivative accounting for the coupling to an AB flux
in the compact $n$-direction.

Technically, Eq. (\ref{eq:12}) represents the main result of the
present paper. We have described the low energy physics of the
quasi-periodic QKR
in terms of the nonlinear $\sigma$-model of disordered
metals~\cite{orth}. The construction is parametrically controlled by
the parameters $\tilde h/K,\omega\ll1 $ and corrections to the
effective action are small in these parameters.  In the following,
we discuss a number of physical predictions deriving from the
representation (\ref{eq:12}).

\textit{Metallic regimes.} In dimensions $d\ge 3$ ($q=\infty$) or
$d\ge 4$ ($q$ finite) the system supports an Anderson
(metal/insulator)
transition. In the metallic phase, $K/\tilde h\gg 1$, fluctuations
are weak and the action may be expanded to quadratic order in the
generators $Z$. Doing the Gaussian integral over $Z$, one then obtains
\begin{align*}
  Y(\varphi,\omega) = {1\over D_\omega \varphi^2-i\omega},
\end{align*}
where $D_\omega\simeq D$ is the diffusion constant weakly
renormalized by non-linear and frequency dependent corrections to the
quadratic theory. Substituting this result into the expression for
$E(t)$, we obtain diffusive growth $E(t)\sim D t$, corresponding to a
finite optical conductivity.

\textit{Localized regimes.} For $q=\infty$, the system is in a
localized phase in low dimensions, $d<3$, or below the Anderson
transition at $K/\tilde h={\cal O}(1)$ in $d\ge 3$. In these regimes,
the diffusion constant is undergoing strong renormalization,
$D_\omega \stackrel{\omega \rightarrow
  0}{\longrightarrow}i\omega$. This in turn leads to saturation $E(t)
\stackrel{t\rightarrow \infty}{\longrightarrow} \mathrm{const.}$,
and vanishing (static) conductivity. For $d\ge 3$, the condition of
scale invariance of the critical conductivity leads to the
predictions $D_\omega\sim (-i\omega)^{d-2 \over d}$ and $E(t) \sim
t^{2\over d}$ at the critical point. This asymptotic agrees with the
experimental observation \cite{Deland10} on the scaling of $E(t)$ in
$d=3$.

\textit{Super-metallic regimes.} For finite $q<\xi$ and $d\le 3$ the
system is localized in the virtual directions and delocalized along
the real angular momentum direction (cf. Fig.  \ref{fig}, c)).
Resonant transmission through the discrete levels of the ensuing
system of effectively finite size then leads to `super-metallic'
growth $E(t) = C t^2$ at large time scales, and a corresponding
diverging (static) conductivity.  The absence of rigorous
descriptions of strong localization notwithstanding (for the
exceptional case of $d=2$ see below), we may apply phenomenological
reasoning to estimate both the coefficient, $C$, and the crossover
time, $t_\xi$, to super-metallic scaling: at short times, the
uncertainty in quasilevel resolution, $\sim t^{-1}$, is larger than
the characteristic quasilevel spacing $\Delta_t\equiv 1/(q
L_t^{d-1})$ of a fictitious system of size $q\times L_t^{d-1}$,
where $L_t\equiv (D_{t^{-1}}t)^{1/2}$ is the characteristic
extension of a diffusive process of duration $t$ in the virtual
directions, and $D_{t^{-1}}$ is a shorthand for the diffusion
coefficient renormalized down to frequency scales $\omega\sim
t^{-1}$.
The level mixing then leads to diffusive growth $E(t)\sim Dt$. The
borderline condition $t^{-1}_\xi = \Delta_{t_\xi}$ marks the
crossover to long-time dynamics, $t>t_\xi$, governed by localization
effects. In this regime, individual
levels are no longer mixed by quantum uncertainty. The coherent
propagation through individual states then leads to $E(t)=Ct^2$,
where the coefficient $C$ is fixed by the matching condition
$D_{t_\xi} t_\xi= C t_\xi^2$, i.e. $C=D_{t_\xi}/t_\xi$.

The quantitative determination of both $\xi$ and $t_\xi$ involves
localization phenomena in slab
geometries and is more difficult. Generally speaking,
the above
condition $q<\xi$ on the transverse extension refers to
the bulk $d$-dimensional localization length.
In contrast, the scale $t_\xi$ is determined
by the localization length
characterizing the quasi $(d-1)$-dimensional low energy regimes. The
application of scaling arguments~\cite{Efetov97} leads to $t_{\xi}
\sim q^2D$ and $t_\xi\sim \exp(2qD)$ in dimensions $d=2$ and $d=3$,
respectively. In dimensions $d>3$, the situation is more complicated
in that the system supports an Anderson
transition in the underlying ($d-1$)-dimensional system. While the
finite transverse extension of the system makes it difficult to
determine the transition point, it is clear as a matter of principle
that lowering the bare value of $D \sim (K/\tilde h)^2$ will trigger
a localization transition, which will manifest itself as a
metal/super-metal transition in the long time scaling of the
observable $E$.

In view of the phenomenological nature of these arguments, it is
reassuring that for $d=2$ a much more sophisticated approach can be
formulated. In this case, the theory is defined on an infinitely long
cylinder, $N=(n,n_1)\in [0,q]\times \Bbb{R}$. Fluctuations of the
field $Q$ inhomogeneous in $n$--direction are penalized by an action
cost (cf. Eq. (\ref{eq:12})) of at least $\sim D/q^2$, which we
identify as the inverse of the diffusion time in $n$-direction.  For
frequency scales smaller than that, $n$-fluctuations can be neglected,
and we are left with the quasi one-dimensional field $Q_N = Q_{n_1}$
with effective action $ S[Q]= -{q\over 8}\int dn_1\, \mathrm{str}
(D(\partial_{n_1} Q )^2 +D \varphi^2
[Q,\sigma^3_\mathrm{AR}]^2+2i\omega \,Q\sigma^3_\mathrm{AR}).  $ The
correlation function $Y(\varphi,\omega)$ of this theory can be
computed by adaption~\cite{at} of Efetov's transfer matrix
technique~\cite{Efetov97}. The qualitative features mentioned above
then follow from scaling properties of the corresponding solutions.
For a quantitative discussion including pre-factors, we refer to
Ref.~\cite{at}.

Summarizing, we have introduced a microscopic theory of quantum phase
transitions in the quasi-periodic QKR. For irrational values of
Planck's constant, the system is described by a $d$--dimensional
nonlinear $\sigma$-model which entails a near perfect analogy to the
physics of $d$--dimensional disordered metals. However, for rational
values, its effective topology changes, and a dimensional reduction to
a quasi $(d-1)$--dimensional system takes place. We discussed the
ensuing consequences, including the existence of a metal/super-metal
quantum phase transition in $d\ge 4$. It stands to reason, that the
interplay of localization and these reduction phenomena should be
observable in experiment.

Work supported by SFB/TR 12, SFB 608 and FOR 960 of the Deutsche
Forschungsgemeinschaft.


\begin{thebibliography}{}

\bibitem{Casati79} B. Chirikov and D. Shepelyansky, Scholarpedia
\textbf{3}, 3550 (2008); S. Fishman, {\it ibid.} \textbf{5}, 9816
(2010).

\bibitem{unit} Throughout we will work in dimensionless units where
  time (kicking strength and Planck's constant) are scaled by (the
  inverse of) the kicking period. The moment of inertia of particle
  is set to unity.

\bibitem{Fishman83} S. Fishman, D. R. Grempel, and R. E. Prange, Phys. Rev. Lett.
\textbf{49}, 509 (1982);
A. Altland, {\it ibid.} \textbf{71}, 69 (1993); A. Altland and M. R.
Zirnbauer, {\it ibid.} \textbf{77}, 4536 (1996); C. Tian, A. Kamenev
and A. Larkin, {\it ibid.} \textbf{93}, 124101 (2004).

\bibitem{Altland10} C. Tian and A. Altland, New J. Phys.
\textbf{12}, 043043 (2010).

\bibitem{Raizen95} F. L. Moore, J. C. Robinson, C. F. Bharucha, Bala Sundaram, and M.
G. Raizen, Phys. Rev. Lett. \textbf{75}, 4598 (1995).

\bibitem{Deland08} J. Chab$\acute{\rm e}$, G. Lemari$\acute{\rm e}$, B. Gr$\acute{\rm e}$aud, D. Delande, P.
Szriftgiser, and J. C. Garreau, Phys. Rev. Lett. \textbf{101},
255702 (2008).

\bibitem{Deland10} G. Lemari$\acute{\rm e}$, H. Lignier, D. Delande, P. Szriftgiser,
and J. C. Garreau, Phys. Rev. Lett. \textbf{105}, 090601 (2010).

\bibitem{Efetov97} K. B. Efetov, {\it Supersymmetry in disorder and chaos} (Cambridge, UK, 1997).


\bibitem{Casati89} G. Casati, I. Guarneri, and D. L. Shepelyansky,
Phys. Rev. Lett. \textbf{62}, 345 (1989).

\bibitem{Smilansky92} R. Bl{\"u}mel and U. Smilansky, Phys. Rev. Lett. \textbf{69}, 217 (1992).

\bibitem{at} C. Tian and A. Altland, in preparation.

\bibitem{Zirnbauer96} M. R. Zirnbauer, J. Phys. A \textbf{29}, 7113
(1996).

\bibitem{orth} More precisely, Eq. (\ref{eq:12}) represents a
  $\sigma$-model of unitary symmetry~\cite{Efetov97}, as relevant for
  systems of broken $\mathrm{T}$-invariance. In the present context,
  $\mathrm{T}$-invariance is broken by the phases $\phi_\pm$. (It gets
  restored after integration over all values of $\phi_\pm$.) At
  off-resonance, i.e., $q\rightarrow \infty$,
the $\mathrm{T}$-breaking is  so weak that we should actually employ the $\mathrm{T}$-invariant form of the QKR nonlinear
  $\sigma$-model (cf. Ref.~\cite{Altland10}). However, this extension is not of great
  consequence and will be discussed elsewhere~\cite{at}.



\end{thebibliography}
\end{document}